\documentclass[%
 aps,
 floatfix,
 amsmath,amssymb,
 reprint,%
 longbibliography,
]{revtex4-1}

\usepackage{graphicx}
\usepackage{dcolumn}
\usepackage{bm}
\usepackage{textgreek}
\usepackage{comment}
\usepackage{soul}
\usepackage{tabularx}
\usepackage{tikz}
\usepackage{textcomp}
\usepackage{multirow}

\usepackage{textcomp}
\usepackage[normalem]{ulem}

\usepackage[labelformat = empty,position=top]{subcaption}
\usepackage[export]{adjustbox}

\newcommand{\mysubfigimg}[3][,]{%
  \setbox1=\hbox{\includegraphics[#1]{#3}}
  \leavevmode\rlap{\usebox1}
  \rlap{\hspace*{1pt}\raisebox{\dimexpr\ht1-1\baselineskip}{#2}}
  \phantom{\usebox1}
}

\begin{document}
\preprint{AIP/123-QED}
\title{Characterization of the $^{12}$C(p,p')$^{12}$C Reaction \\ ($E_p$=19.5--30~MeV) for Active Interrogation}
\author{J.~Nattress$^{1,2}$}
\author{F.~Sutanto$^{1}$}
\author{P.-W.~Fang$^{3}$}
\author{Y.-Z.~Chen$^{3}$}
\author{A. Cheng$^{3}$}
\author{K.-Y.~Chu$^{4}$}
\author{T.-S.~Duh$^{4}$}
\author{H.-Y.~Tsai$^{3}$}
\author{M.-W.~Lin$^{3}$}
\email{mwlin@mx.nthu.edu.tw}
\author{I.~Jovanovic$^{1}$}

\affiliation{$^1$Department of Nuclear Engineering and Radiological Sciences, University~of~Michigan, Ann~Arbor, MI~48109~USA}
\affiliation{$^2$Oak Ridge National Laboratory, Oak Ridge, TN 37830 USA}
\affiliation{$^3$Institute of Nuclear Engineering and Science, National Tsing Hua University, Hsinchu 30013, Taiwan}
\affiliation{$^4$Isotope Application Division, Institute of Nuclear Energy Research, Taoyuan City 32546, Taiwan}

\date{\today}
\begin{abstract}
\noindent Passive detection of special nuclear material (SNM) is challenging due to its inherently low rate of spontaneous emission of penetrating radiation, the relative ease of shielding, and the fluctuating and frequently overwhelming background. Active interrogation, the use of external radiation to increase the emission rate of characteristic radiation from SNM, has long been considered to be a promising method to overcome those challenges. Current active interrogation systems that incorporate radiography tend to use bremsstrahlung beams, which can deliver high radiation doses. Low-energy ion-driven nuclear reactions that produce multiple monoenergetic photons may be used as an alternative. The $^{12}$C(p,p')$^{12}$C is one such reaction that could produce large yields of highly penetrating 4.4- and 15.1-MeV gamma rays. This reaction does not directly produce neutrons below the $\sim$19.7-MeV
threshold, and the 15.1-MeV gamma-ray line is well matched to the photofission cross section of $^{235}$U and $^{238}$U. We report the measurements of thick-target gamma-ray yields at 4.4 and 15.1~MeV from the $^{12}$C(p,p')$^{12}$C reaction at proton energies of 19.5, 25, and 30~MeV. Measurements were made with two 3'' EJ309 cylindrical liquid scintillation detectors and thermoluminescent dosimeters placed at 0\textdegree~and 90\textdegree, with an additional 1.5'' NaI(Tl) cylindrical scintillation detector at 0\textdegree. We estimate the highest yields of the 4.4- and 15.1-MeV gamma rays of 1.65$\times10^{10}$ sr$^{-1}$\textmu C$^{-1}$ and 4.47$\times10^8$ sr$^{-1}$\textmu C$^{-1}$ at a proton energy of 30~MeV, respectively. The yields in all experimental configurations are greater than in a comparable deuteron-driven reaction that produces the same gamma-ray energies -- $^{11}$B(d,n$\gamma$)$^{12}$C. However, a significant increase of the neutron radiation dose accompanies the proton energy increase from 19.5 to 30~MeV. 

\end{abstract}
%
\maketitle

\section{Introduction}
Screening and interdiction of shielded special nuclear material (SNM) is a significant nuclear security concern. The problem encompasses not only cargo containers but also smaller objects, as well as the movement of illicit materials into and out of enrichment facilities~\cite{kouzes2005detecting,runkle2009photon}. Several million ocean cargo containers arrive at United States seaports annually~\cite{grassleycontainer,thibault2006response} with over half the distribution concentrated at three ports -- Los Angeles, Long Beach, and New York~\cite{grassleycontainer}. The large traffic and its uneven distribution raise serious concerns about the number of containers that can be accurately scanned without impeding commerce and still maintaining nuclear security vigilance.

The technique of active interrogation has been of increasing interest in recent years for detecting SNM~\cite{jovanovic2018active}. This is based on an expectation that the increased magnitude and the time structure of the fission signatures imposed by active interrogation can increase the probability of SNM detection. New and innovative approaches are needed to overcome specific application constraints, such as the maximum scanning time and radiation dose. Traditional active interrogation systems tend to employ energetic X-ray transmission radiography using continuous-energy-spectrum bremsstrahlung beams. Such beams can deliver high radiation doses, especially at high electron energies needed to produce sufficiently energetic photons for penetrating radiography and inducing photofission. The lower-energy photons in the bremsstrahlung spectrum, well below the photofission threshold, have poor penetration through objects with high areal density and may significantly contribute to the imparted radiation dose while not inducing fission~\cite{geddes2017impact,jones2009bremsstrahlung}. As a result, an alternative method for production of quasi-monoenergetic, continuously-tunable high-energy X-rays that has garnered considerable attention in the recent period is laser Compton scattering~\cite{Liu:2014}.

Low-energy ion-driven nuclear reactions that produce multiple monoenergetic, high-energy gamma rays could be used as an alternative to bremsstrahlung in active interrogation applications. Reactions that exhibit a large positive $Q$-value allow the use of lower-energy (and thus typically smaller and simpler) accelerators to create highly penetrating energetic photons. Such monoenergetic or multiple-monoenergetic sources may offer benefits such as improved performance in radiography and higher fission rates per unit of imparted radiation dose. Near-monoenergetic photon sources at or near the photofission resonance ($\sim$15~MeV) offer a factor of 4 lower imparted radiation dose when compared to a bremsstrahlung source with a 19-MeV endpoint energy~\cite{geddes2017impact}. Low-energy nuclear reactions, therefore, represent a promising method to produce quasi-monoenergetic photon beams which could lead to construction of lower-dose, small-footprint systems to detect illicit movement of SNM as well as support a broad range of measurements in the area of nuclear security, nuclear nonproliferation, and nuclear safety.

\begin{figure}[h!]
	\centering
	\includegraphics[width=0.45\textwidth]{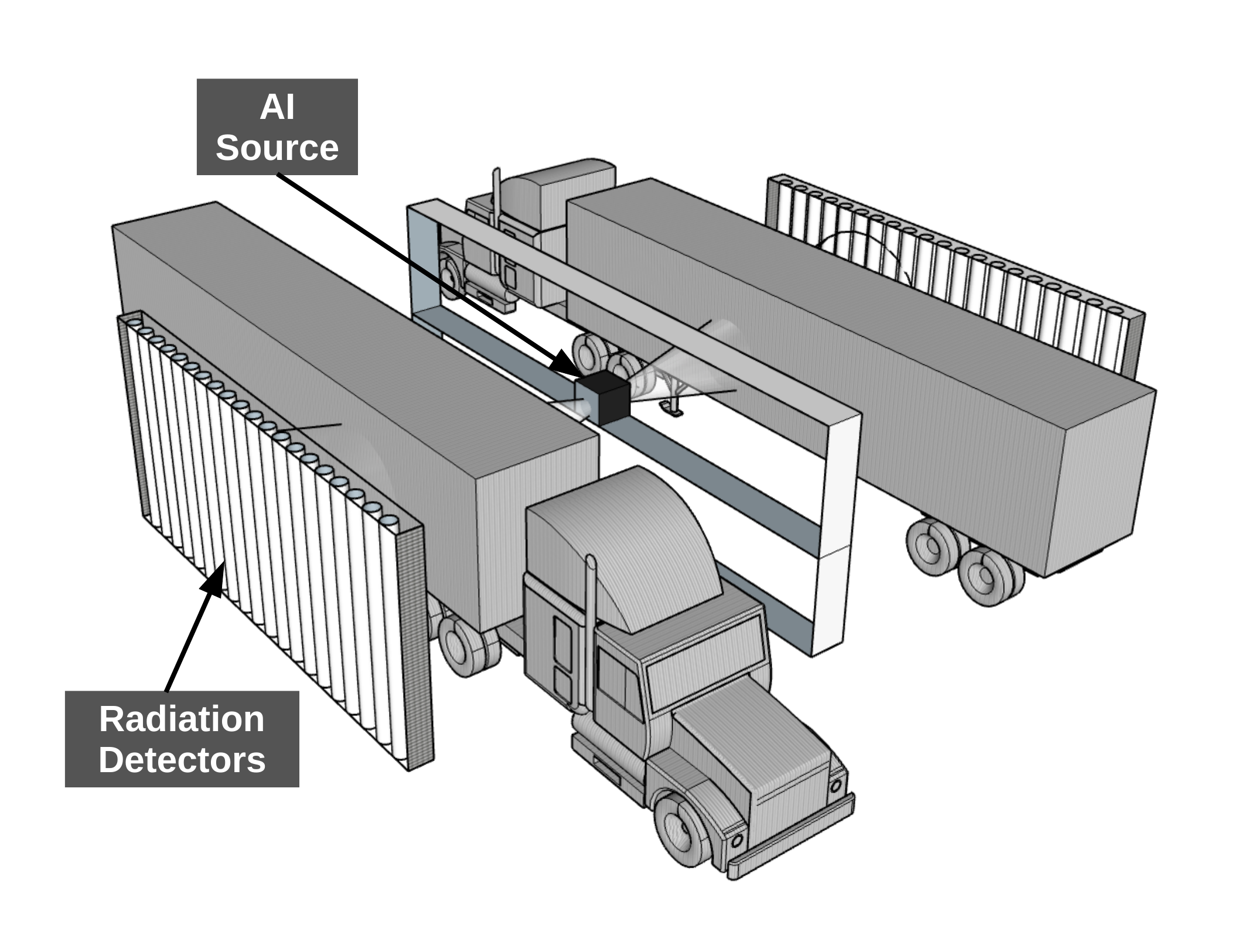}
	\caption{Conceptual design of a cargo scanning system using a low-energy nuclear reaction based active interrogation source. A single source may be used to implement simultaneous scanning of multiple cargo streams~\cite{nattress2018discriminating}.}
	\label{fig:system}
\end{figure}

Figure~\ref{fig:system} illustrates a conceptual design of one specific application -- a cargo screening active interrogation system -- where a low-energy nuclear reaction source could be used to increase the scanning rate by implementing simultaneous scanning of multiple cargo streams. Such a system would need to address some of the major specifications for cargo container scanning, including an acceptable radiation dose imparted to the cargo, penetration capability, container scanning speed, and material discrimination performance~\cite{worlCustoms}.

Several candidate reactions have been previously identified that produce prolific yields of energetic gamma rays, including those above the photofission energy threshold, or neutrons that may also be viable for active interrogation~\cite{taddeucci2007neutron}. Such high-energy particles can be used to penetrate dense cargo to perform transmission radiography for material identification~\cite{Rose2016,henderson2018experimental,nattress2019} or induce fission~\cite{Mayer2016,nattress2018}. In Ref.~\cite{taddeucci2007neutron} several candidate reactions were considered, including (p,$\gamma$), (p,$\alpha\gamma$), and (d,n$\gamma$). Four candidate targets ($^{11}$B, $^7$Li, $^{19}$F, and $^{15}$N) were selected based on the reaction \textit{Q}-values and gamma-ray energies from the daughter excited nuclei. Protons, deuterons, and tritons were used as projectiles. The $^{11}$B(d,n)$^{12}$C reaction ($Q$=13.6~MeV) has emerged as one of the leading candidates for an active interrogation source~\cite{taddeucci2011high}. With deuteron energies of several MeV or higher, this reaction produces a strong 15.1-MeV gamma-ray energy line that is well overlapped with the photofission cross section of $^{235}$U and $^{238}$U~\cite{chadwick2006endf}. This reaction, however, has a high neutron yield, approximately 20--50 times the relevant gamma-ray yield~\cite{Rose2016}. 

With the continued advancement of superconducting cyclotron technology, more compact, deployable active interrogation systems utilizing nuclear reactions that do not directly produce neutrons can be envisioned. One such reaction which can excite the 15.1-MeV state of $^{12}$C inelastically is $^{12}$C(p,p')$^{12}$C~\cite{ajzernberg:1990,warburton:1962,measday:1963,howell:1980}. This reaction has an incident proton threshold energy of 16.39~MeV for exciting the 15.1-MeV $^{12}$C state and does not directly produce neutrons below the $\sim$19.6-MeV threshold~\cite{rimmer1968resonances, berghofer1976high}. Gamma rays at 4.4~MeV are also produced prolifically in this reaction, and can be used in conjunction with 15.1-MeV gamma rays to perform spectroscopic transmission radiography and identify the atomic number of an unknown material~\cite{Rose2016,henderson2018experimental}.  

Organic scintillation detectors are typically used to perform fast neutron spectroscopy, where pulse-shape analysis enables the discrimination between neutrons and photons. While it is well-known that low-\textit{Z} organic scintillators are also sensitive to photons and their response is nearly proportional, photon spectroscopy is usually not attempted due to low efficiency, resolution, and the absence of photopeak features in their light-output spectrum. Although the presence of prominent full-energy deposition peaks is often favored for photon spectroscopy, many common high-\textit{Z} scintillators depend on expensive crystalline materials with slow decay timing characteristics, which perform poorly in high-rate environments. They can also be costly to scale to large volumes, or may not be able to distinguish between neutron and photon interactions. The fast rise-time and decay-time response of organic liquid scintillator, along with the ability to detect and identify both neutrons and photons, makes it an attractive choice for an active interrogation system that requires simultaneous photon and neutron identification in a mixed-radiation, high-rate environment. If the photon source has a spectrum that features well-separated discrete energies, discernible spectral features can be observed in the resulting light-output spectrum, which can be exploited to perform spectroscopy~\cite{nattress2017response}. 

We present gamma-ray yield measurements of the $^{12}$C(p,p')$^{12}$C reaction made with two organic liquid scintillators placed at 0\textdegree~and 90\textdegree~with respect to the proton beam axis, with an additional Na(Tl) detector placed at 0\textdegree~for experimental validation. We further report neutron and gamma-ray dose measurements. We observe an increase in the production rate of gamma rays with an increase in proton energy, as well as a rapid increase in the measured neutron radiation dose with proton energy. The active interrogation source proposed in this work focuses on the key performance parameters -- the emitted radiation flux, energy spectrum, and dose rate. Other essential design considerations such as radiation shielding requirements must be considered before final implementation. 
\section{Materials and Methods}
The experiments were conducted using a cyclotron located at the Institute of Nuclear Energy Research in Taiwan. This accelerator produces proton pulses at a fixed frequency of 73.2~MHz, and the proton energy can be adjusted from 15 to 30~MeV with a maximum average current of 10~\textmu A delivered to the experimental station. A 6.35-mm-thick natural carbon target consisting of 98.9\% $^{12}$C and 1.1\% $^{13}$C and an active area of 2~cm $\times$ 2~cm was held by an aluminum mount and placed at the center of an aluminum vacuum chamber (20~cm $\times$ 20~cm $\times$ 20~cm, 1.2~cm-thick). To prevent protons from striking the target mount and generating neutrons at the target location, a 1-cm-thick aluminum collimator was inserted into the upstream beamline for shaping the incident proton beam to a transverse diameter of 1~cm before entering the target chamber. This collimation, however, inevitably produces additional neutrons when protons in the outer radial region of each pulse strike the aluminum collimator. Therefore, a water tank (1~m $\times$ 1~m $\times$ 1~m) was placed to surround the vacuum pipe for shielding neutrons emitted from the collimator. The on-target proton current was calibrated when the target mount was replaced by a Faraday cup, whose output was connected to a digital picoammeter (Keithley, model 486) with a resolution of 10~fA. 

\begin{figure}[!h]
\centering
\begin{minipage}{0.5\textwidth}
\includegraphics[width=8cm]{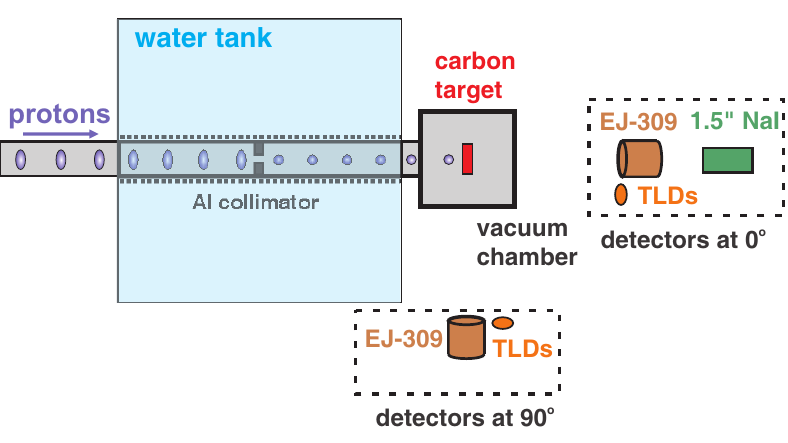}
\end{minipage}
\caption{Schematic of the experimental setup}
\label{Fig:setup}
\end{figure}

The radiation produced by the source was measured with one NaI(Tl) scintillation detector, two EJ309 organic liquid scintillation detectors~\cite{ej309}, and pairs of thermoluminescent dosimeters (TLDs). The NaI(Tl) detector was placed at 0$^{\circ}$ at a distance of 2.05~m from the carbon target, while the EJ309 liquid scintillation detectors and TLDs were placed at 0$^{\circ}$ and 90$^{\circ}$, at the same distance of 1.4~m. Here, the angles are given with respect to the propagation axis of protons incident on the target. The NaI(Tl) dimensions were 1.5'' (diameter) $\times$ 1.5'' (height); the EJ309 detectors were 3'' in diameter $\times$ 3'' in height. Figure~\ref{Fig:setup} shows a diagram of the experimental setup including the aluminum collimator, vacuum target chamber, water tank for neutron shielding, one NaI(Tl) scintillation detector, two EJ309 liquid scintillators, and two paired TLDs.
Table~\ref{table1} summarizes the positions and solid angles subtended by detectors, as measured relative to the target. The detectors were placed at nearly the same heights as the target height (1.5~m above the ground level).
\begin{table}[!h]
\caption{\label{table1} Summary of characteristics of the scintillation detectors and their placement relative to the target}
\begin{ruledtabular}
\begin{tabular}{*{5}{c}}
Angle &\multicolumn{1}{c}{Detector} & \multicolumn{1}{c}{Radial} & \multicolumn{1}{c}{Height} & \multicolumn{1}{c}{Solid} \\ 
&\multicolumn{1}{c}{} & \multicolumn{1}{c}{Distance (m)} & \multicolumn{1}{c}{(m)} & \multicolumn{1}{c}{Angle (sr)} \\ 
\cline{1-5}
\multirow{2}{*}{0$^{\circ}$} &\multicolumn{1}{c}{1.5''$\times$1.5'' NaI(Tl)}& \multicolumn{1}{c}{2.05}& \multicolumn{1}{c}{1.53}& \multicolumn{1}{c}{2.7$\times$10$^{-4}$}\\
&\multicolumn{1}{c}{3''$\times$3'' EJ309}& \multicolumn{1}{c}{1.4}& \multicolumn{1}{c}{1.45}& \multicolumn{1}{c}{2.3$\times$10$^{-3}$}\\
\hline 
90$^{\circ}$
&\multicolumn{1}{c}{3'' $\times$ 3'' EJ309}& \multicolumn{1}{c}{1.4}& \multicolumn{1}{c}{1.2}& \multicolumn{1}{c}{2.3$\times$10$^{-3}$}\\ 
\end{tabular}
\end{ruledtabular}
\end{table}
Three proton-projectile-target configurations were tested. The compact superconducting cyclotron was operated at 0.35~nA, 0.3~nA, and 0.33~nA on-target current with 19.5~MeV, 25~MeV, and 30~MeV proton energies incident onto the carbon target, respectively. At each proton energy data were recorded for 8~hours. 

\begin{figure}
  \centering
  \begin{tabular}{@{}p{1\linewidth}@{}p{1\linewidth}@{}}
    \mysubfigimg[width=\linewidth]{(a)}{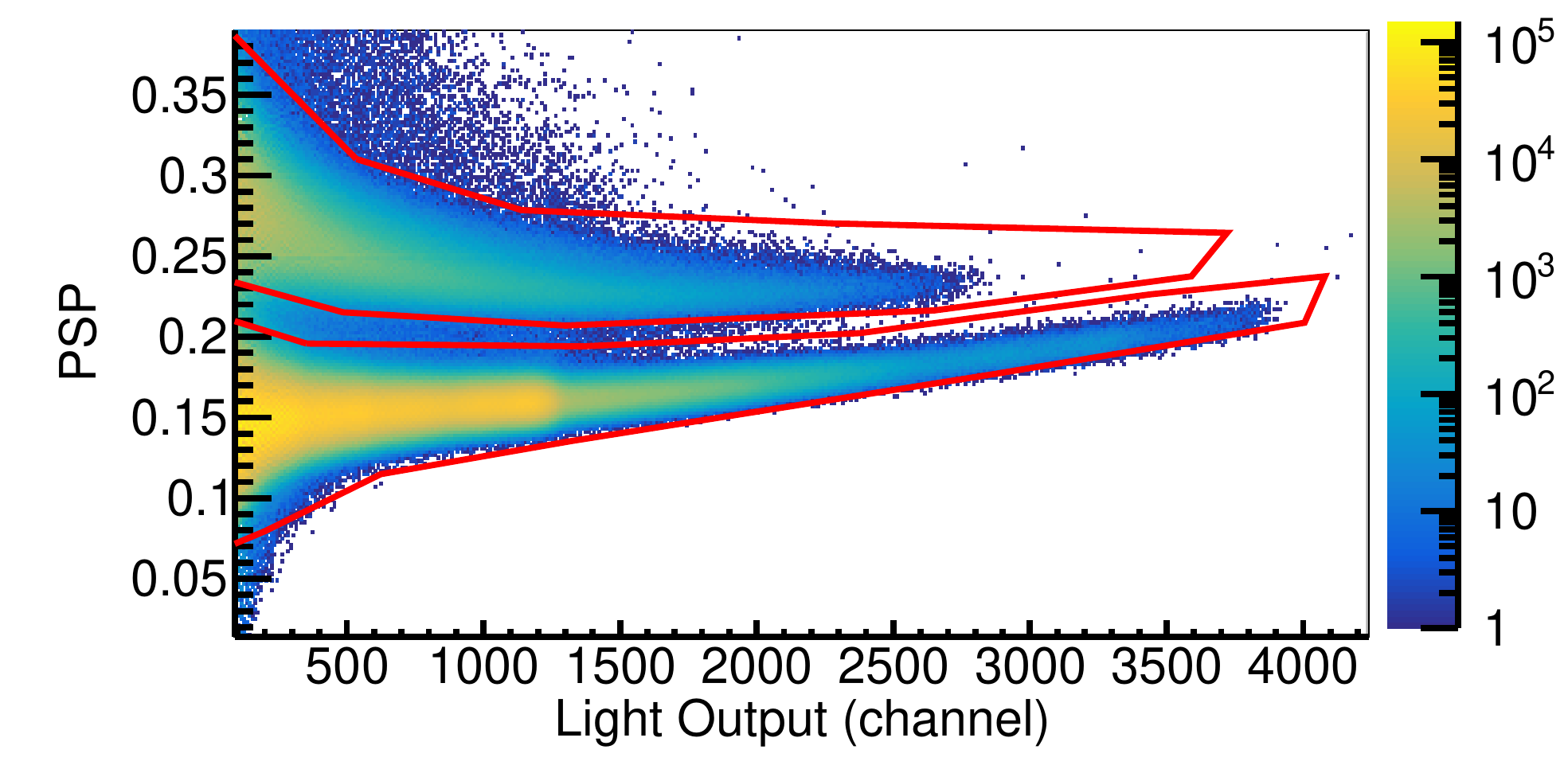} \\ 
    \mysubfigimg[width=\linewidth]{(b)}{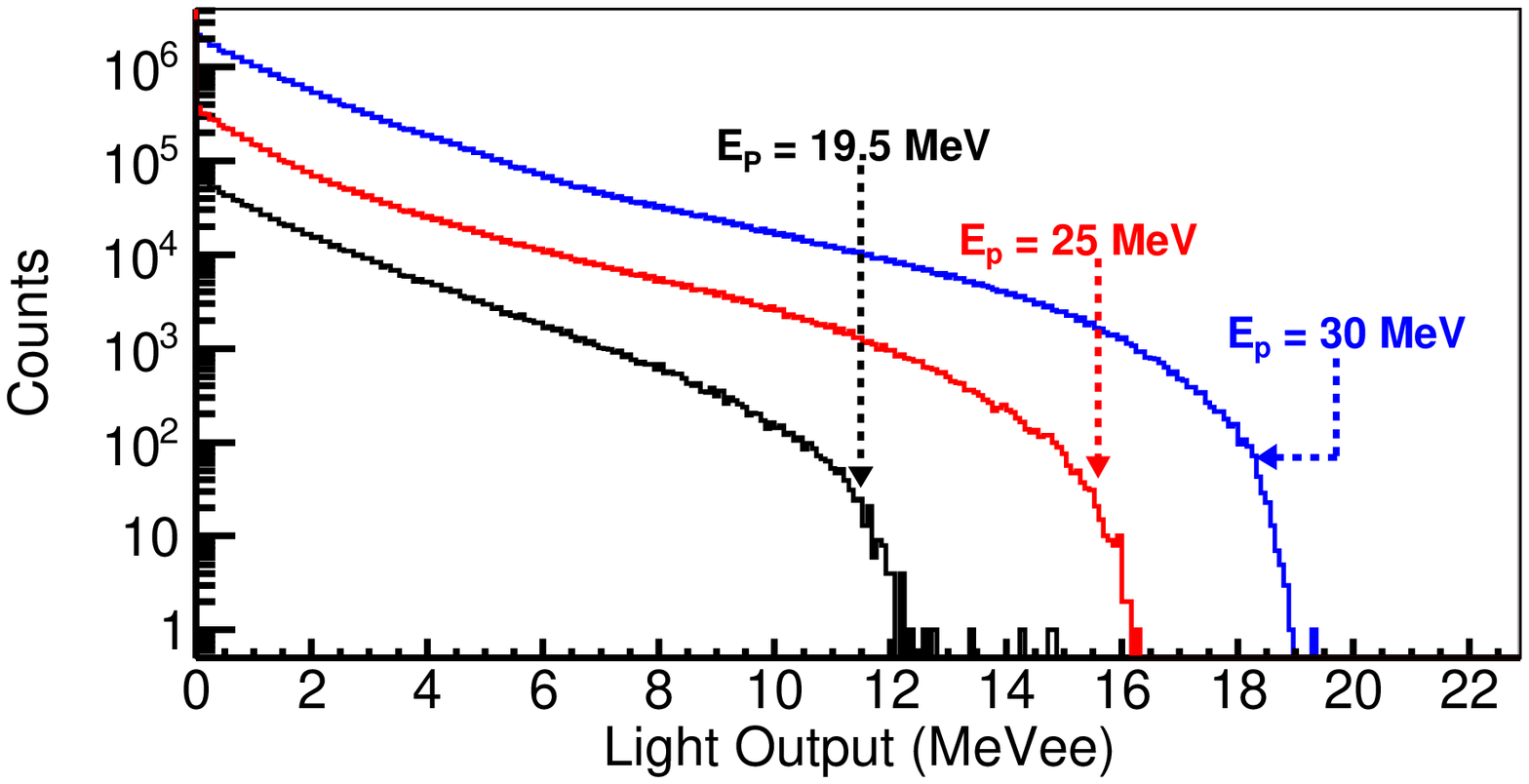} \\ 
  \end{tabular}
  \caption{(a)~Response of a 3'' EJ309 detector for 19.5-MeV protons with the neutron (higher $PSP$) and photon (lower $PSP$) fiducial cuts shown in red and (b)~the neutron light-output response for 19.5-, 25-, and 30-MeV protons incident on a natural carbon target for an 8-hour measurement time.}
 \label{fig:LOvsPSP}
\end{figure}

\begin{figure}[]
\vspace{-.5em}
\centering
\includegraphics[width=\linewidth]{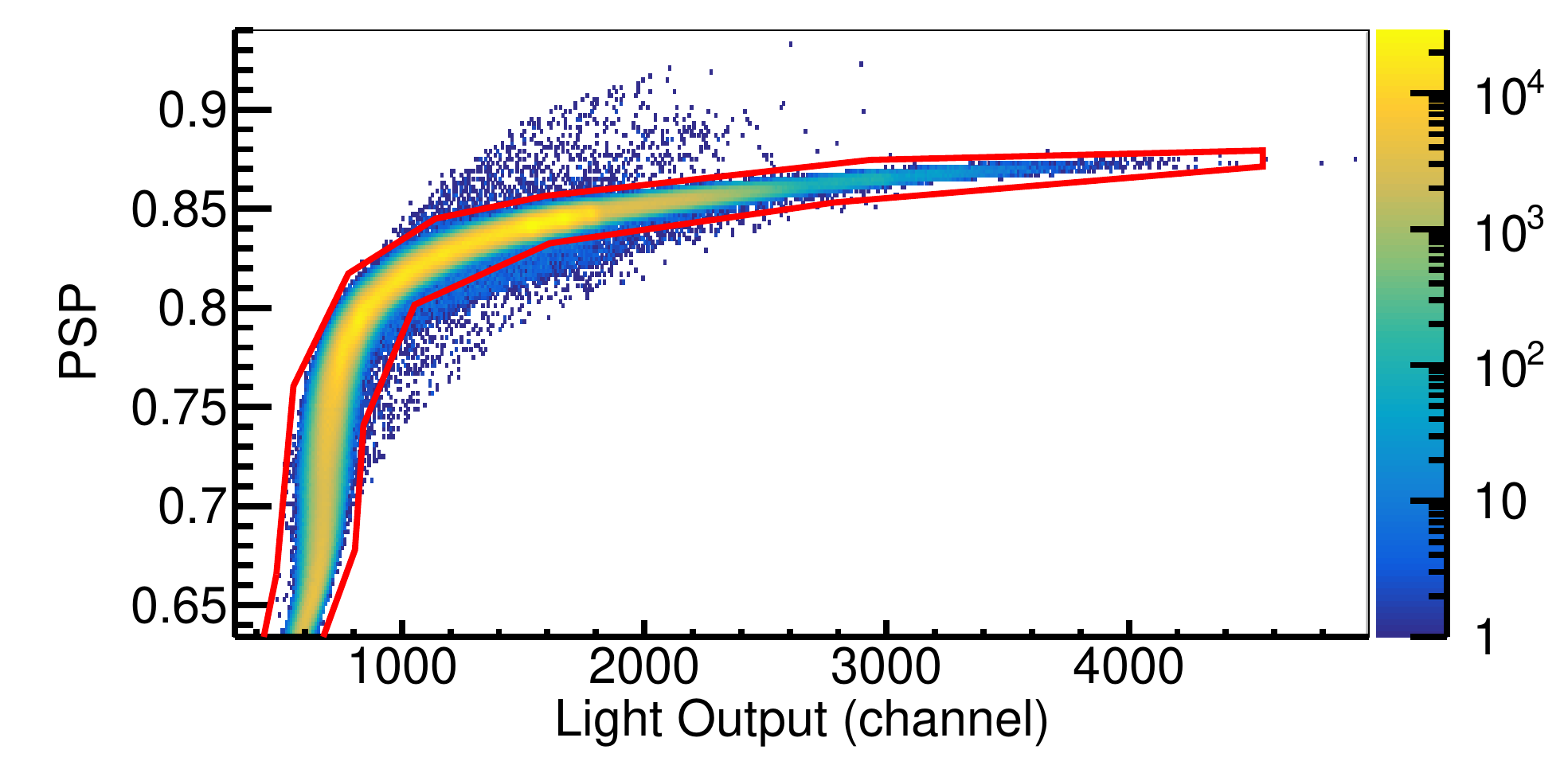}
\caption{Response of a 1.5'' NaI(Tl) detector for 19.5-MeV protons incident on a natural carbon target for an 8-hour measurement time. The photon fiducial cut is shown in red.}\label{fig:naPSP}
\vspace{-.5em}
\end{figure}

Signals generated by the scintillation detectors were digitized using an 8-channel, 14-bit, 500-MHz waveform digitizer (CAEN DT5730) with a 992-ns long waveform record. The CAEN Multi-PArameter Spectroscopy Software (CoMPASS)~\cite{CAENCoMPASS} was used to record the integrated charge in two distinct time windows measured with respect to the pulse trigger. We refer to these two recorded quantities as the long-gate integral ($Q_{long}$) and the short-gate integral ($Q_{short}$). Pulse-shape analysis was performed using the digitizer field-programmable gate array by calculating the pulse shape parameter ($PSP$) for each signal waveform, which is defined as
\begin{equation}\label{PSP}
PSP = \left(Q_{long}-Q_{short}\right)/Q_{long}.
\end{equation}
For the NaI(Tl) detectors, $Q_{long}$ and $Q_{short}$ were defined as the total charge in the time windows [$t_s$, $t_s$+400~ns] and [$t_s$, $t_s$+100~ns], respectively, with the reference delay $t_s=50$~ns after the leading edge trigger. The time windows for the EJ309 detectors were [$t_s$+400~ns] and [$t_s$+54~ns] for $Q_{long}$ and $Q_{short}$, respectively, where the trigger delay was set to $t_s=10$~ns. By performing the fiducial cut in the PSP-light output parameter space of the EJ309 response, single-photon events can be distinguished from both neutron and pile-up events. A PSP-light output scatter plot with the fiducial cuts (red) for one of the EJ309 detectors and the NaI(Tl) detector is shown in Fig.~\ref{fig:LOvsPSP} and Fig.~\ref{fig:naPSP}. Also in Fig.~\ref{fig:LOvsPSP}, the neutron light-output response for 19.5-, 25-, and 30-MeV protons incident on a natural carbon target is shown, indicating an increase in neutron energy with increasing incident proton energy. Table~\ref{table:data_rate} summarizes the observed event rates over the 8-hour measurement time. The 3-in EJ309 detector at 0\textdegree exhibits the highest event rate in each experiment. Meanwhile, EJ309 at 90\textdegree and NaI(Tl) at 0\textdegree experience lower event rates ($\sim$60\% and $\sim$10\% of the EJ309 detector at 0\textdegree, respectively). According to the timestamps recorded by the digitizer, the dead time of the detection system is calculated to be $\approx$1.024 \textmu s; therefore, the true counting rates are estimated according to the non-paralyzable model~\cite{Knoll_RDM} and are listed in Table~\ref{table:data_rate}. 
\begin{table}[h]
\vspace{.5em}
\caption{Summary of the observed and the estimated true counting rates.}
\label{table:data_rate}
\centering
\begin{tabular}{cccc}
\hline
\hline
Detector  &      Proton    energy   &   \multicolumn{2}{c}{Counting rate  }  \\
          &      (MeV)              &  Observed (kHz)       &  True  (kHz)   \\
\hline
\multirow{3}*{EJ309-0\textdegree}    & 19.5   &   2.823     &  2.831     \\
                                     & 25     &   7.6       &  7.656     \\
                                     & 30     &   27.013    &  27.782    \\ 
\hline
\multirow{3}*{EJ309-90\textdegree }  & 19.5   &   1.92      & 1.923      \\
                                     & 25     &   4.902     &  4.927     \\
                                     & 30     &   15.821    &  16.082    \\
\hline
\multirow{3}*{NaI-0\textdegree}      & 19.5   &   0.376     &   0.376    \\
                                     & 25     &   0.943     &   0.944    \\
                                     & 30     &   2.81      &  2.818     \\
\hline
\end{tabular}	
\vspace{-.5em}
\end{table}

%
%

%

\begin{figure}[]
\vspace{-.5em}
\centering
\includegraphics[width=\linewidth]{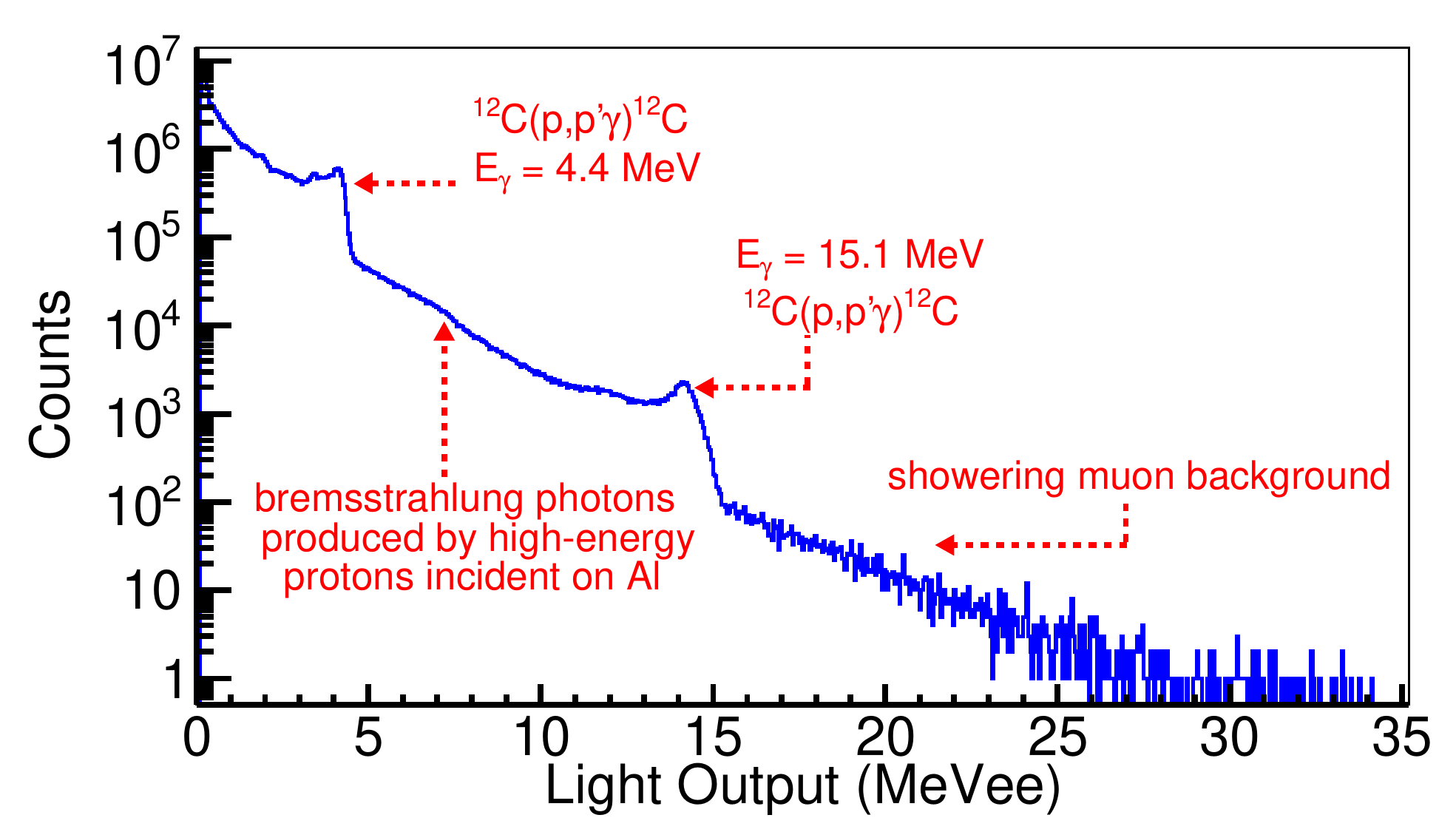}
\caption{Recorded light-output spectra measured with a 3'' EJ309 detector detector at 90\textdegree~over a measurement time of 8~hours for a proton energy of 25~MeV incident on a natural carbon target.\label{Fig:gammaLO}}
\vspace{-.5em}
\end{figure}

 Fiducial cuts were used to select photon and reject pileup events, and events within the respective cuts were used to produce the photon light-output distributions for each detector for three proton-projectile-target experimental configurations. Figure~\ref{Fig:gammaLO} shows a representative photon light-output distribution which corresponds to the EJ309 detector response at 90\textdegree~at a proton energy of 25~MeV. 
In Fig.~\ref{Fig:gammaLO} the double-escape peaks from 4.4-MeV and 15.1-MeV gamma rays are readily observable, along with the Compton-scattering feature at 4.4~MeV~\cite{nattress2017response}. As expected, there is no observable photopeak due to the low atomic numbers of liquid scintillation detector's constituents. The broad energy distribution at high light output ($\gtrsim$20~MeVee) is attributed to muon energy loss in the detector volume. The continuum between the 4.4-MeV and 15.1-MeV spectral features is attributed to bremsstrahlung produced by high-energy protons incident on the aluminum collimator placed upstream from the target; this is corroborated by simulation as discussed in Sect.~III.  

Photon and neutron radiation doses at 0\textdegree~and 90\textdegree~were measured for 8 hours and distinguished from each other based on the dual-TLD method~\cite{Hsu2008}. In this method, pairs of TLD~600 and TLD~700 chips (3.2 $\times$ 3.2 $\times$ 0.89~mm, Harshaw) were placed at the distance 1.4~m away from the carbon target at each angle. The TLD~700 is primarily sensitive to photons, since 99.993\% of its lithium isotope content is $^{7}$Li, which exhibits a much lower neutron response than that of $^{6}$Li. In contrast, the 95.62\% enrichment of $^{6}$Li in TLD~600 allows it to record the dose contributed from both neutrons and photons. By placing paired TLDs in a mixed photon-neutron field, the TLD~600 reading $Q_6$ accounts for the majority of neutron response reading $Q_{6,n}$ and the minority of photon response reading $Q_{6,p}$; while the TLD~700 reading $Q_7$ is dominated by the photon response reading $Q_{7,p}$ when compared to the neutron response reading $Q_{7,n}$.

In our experiments, the total reading $Q_6$ and $Q_7$ measured at a given proton energy is averaged from three TLD chips of each type placed at each angle, which also allows for consistency checks. Note that, when TLD~600 and TLD~700 are placed at the same position, the measured $Q_{6,p}$ and $Q_{7,p}$ represent the same photon dose. This photon dose $D_{ph}$ can be calibrated as
\begin{equation}\label{g_calib}
  D_{ph}=Q_{x,p}\times CF_{x,p},
\end{equation}
where $CF_{x,p} $ stands for the calibration factor and $x=\{6,7\}$ relates to TLD 600 and TLD 700, respectively. These calibration factors were measured to be $CF_{6,p}\sim 1.8 \times 10^{-5}$~mGy/nC and $CF_{7,p}\sim 1.6\times10^{-5}$~mGy/nC using the $^{137}$Cs field established in a dedicated calibration setup. Once the photon dose $D_{ph}$ is determined from the TLD~700 measurements, the photon response $Q_{6,p}$ of TLD~600 included in the total reading $Q_{6}$ can be estimated by
\begin{equation}\label{6_g_read}
Q_{6,p}=D_{ph}/CF_{6,p}, 
\end{equation} 
since TLD~600 experiences the same $D_{ph}$ as the paired TLD~700. As a result, the neutron response reading $Q_{6,n}$ of the TLD~600 is determined from
\begin{equation}\label{6_n_read}
Q_{6,n} = Q_{6} - Q_{6,p}.
\end{equation}
The TLD~600 chips were calibrated using a $^{252}$Cf source, yielding a factor $CF_{H,n}\sim 1.7 \times10^{-4}$~mSv/nC, which relates the neutron response $Q_{6,n}$ to the neutron dose equivalent $H_{n}$ (mSv):
\begin{equation}\label{n_H_calib} 
  H_{n}=CF_{H,n}\times Q_{6,n}.
 \end{equation}
 %

\section{Light-output Spectrum Simulation}
Monte Carlo simulations were performed using Geant4.10.3~\cite{geant4:2003} to help assess the spectral features in the recorded photon light-output distributions. The physics list QGSP\_BERT\_HP was used to model the particle interactions. The model includes the carbon target, vacuum chamber, collimator, water tank, and the surrounding concrete wall; in this way, the effect of shielding and the production of bremsstrahlung can be investigated and applied when analyzing the experimental results. 

To investigate the continuum observed in the 5--10~MeV energy range, a simulation of high-energy proton interactions in a 1-cm-thick aluminum slab was performed. Bremsstrahlung photons propagate through the water tank and their energies are tallied. The softened spectrum is then used to sample the energy of a new group of photons, for which the direction of propagation is biased towards the detector. The simulated detector response to bremsstrahlung is obtained and smeared with the resolution function. The resolution is parameterized as outlined in Ref.~\cite{nattress2017response}. Monte Carlo simulations of high-energy gamma-ray (4.4 and 15.1~MeV) interactions in the EJ309 detectors have also been performed to obtain the responses of individual detectors used in the experiments. 

To calculate the gamma-ray yields at 4.4 and 15.1~MeV, four individual spectra -- measured background, simulated response to bremsstrahlung at the respective proton energy, and the simulated responses to 4.4 and 15.1~MeV photons -- are fit to the measured data. A $\chi^2$-analysis was used to determine the parameters that maximize the agreement between the measured and the simulated detector responses. Local function minimization was performed using MIGRAD, HESSE, and MINOS algorithms as implemented in the ROOT framework~\cite{brun:1997}.
This analysis was conducted for each detector at each of the three proton energies. The resulting fit spectra were used to calculate the yields of 4.4- and 15.1-MeV gamma rays normalized to the solid angle and measured cyclotron current. The gamma-ray yields measured with the EJ309 detectors are reported in Tables~\ref{table:44} (4.4~MeV) and~\ref{table:15} (15.1~MeV), with the NaI(Tl) detector producing results similar to the EJ309 detector at 0\textdegree. 

\begin{figure}
  \centering
  \begin{tabular}{@{}p{1\linewidth}@{}p{1\linewidth}@{}}
    \mysubfigimg[width=\linewidth]{(a)}{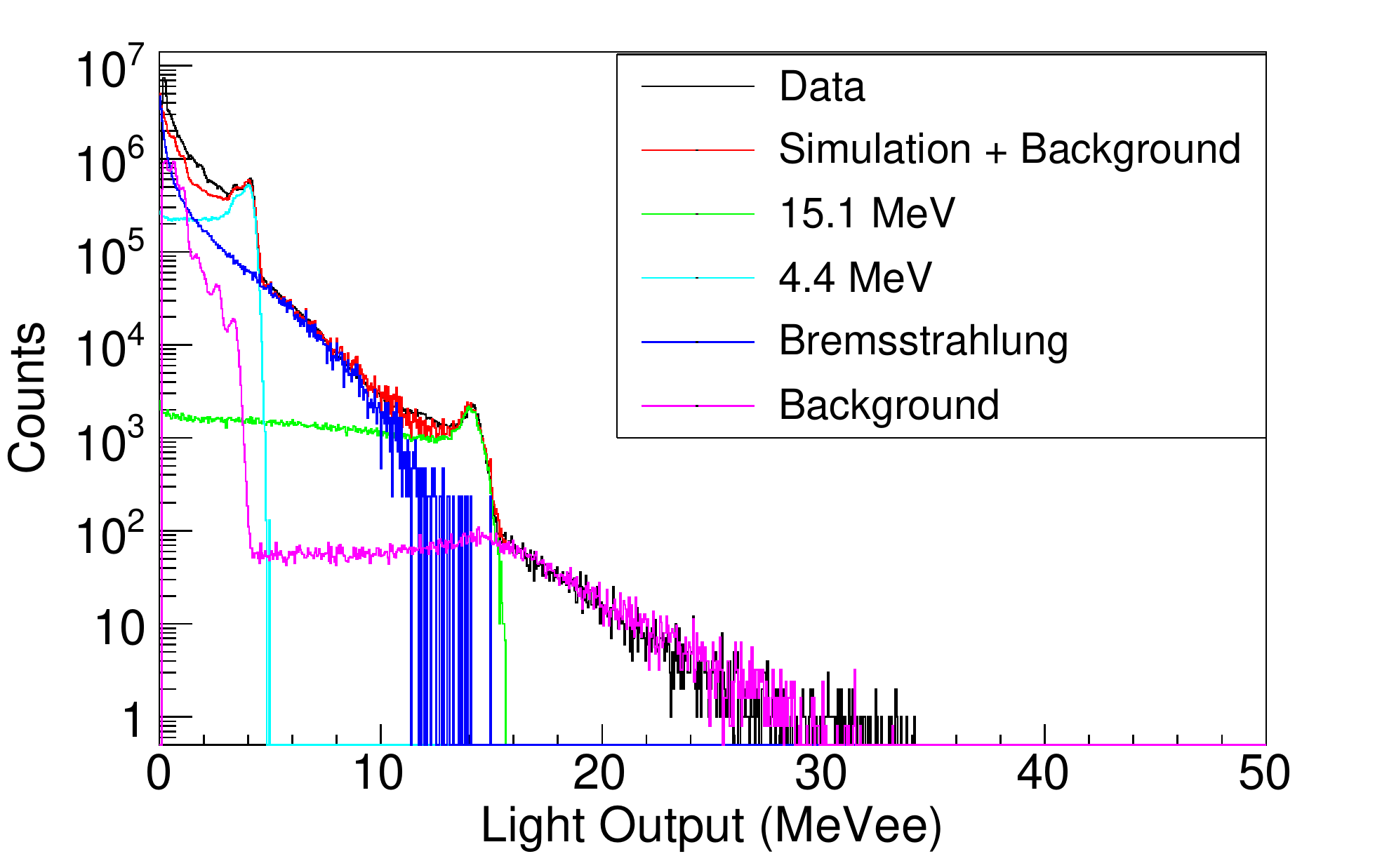} \\ 
    \mysubfigimg[width=\linewidth]{(b)}{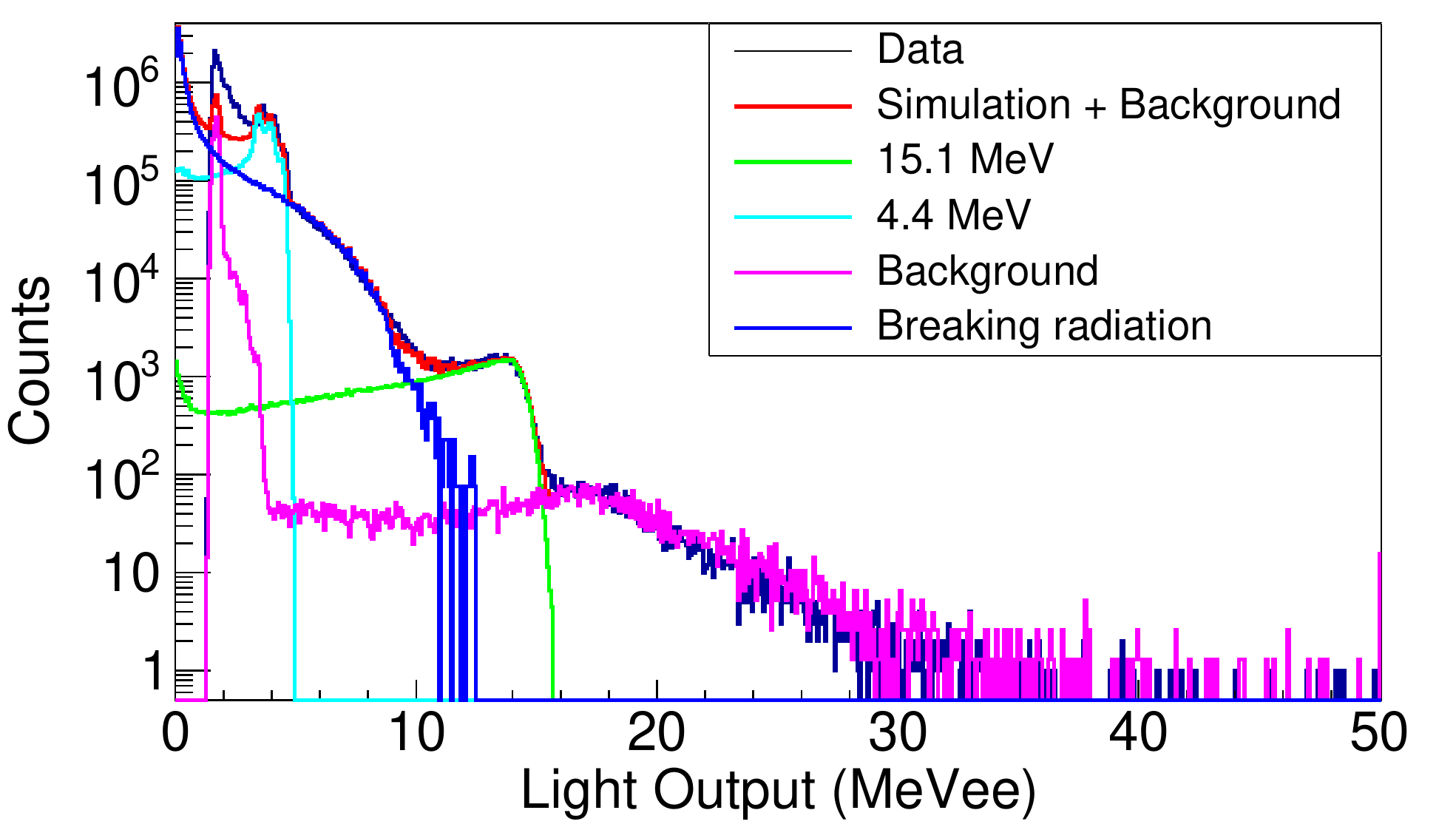} \\ 
  \end{tabular}
  \caption{Experimental light output and simulated contributions to light output for (a)~3'' EJ309 detector detector at 90\textdegree~and (b) 1.5'' NaI(Tl) detector detector at 0\textdegree~for a 25-MeV proton incident onto a thick natural carbon target over a measurement time of 8~hours.}
 \label{fig:indCont}
\end{figure}

\begin{figure}
  \centering
  \begin{tabular}{@{}p{1\linewidth}@{}p{1\linewidth}@{}}
    \mysubfigimg[width=\linewidth]{(a)}{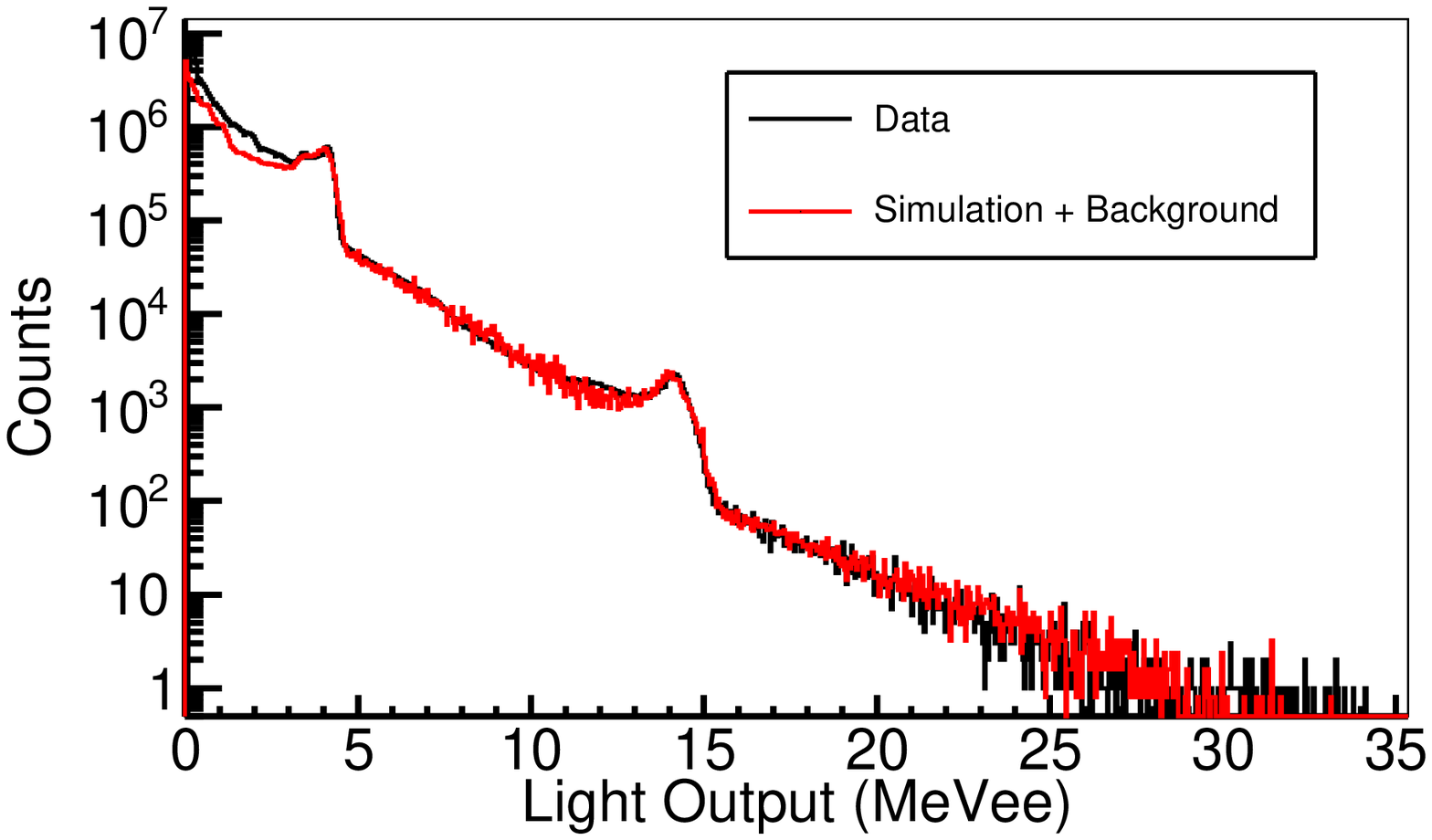} \\
    \mysubfigimg[width=\linewidth]{(b)}{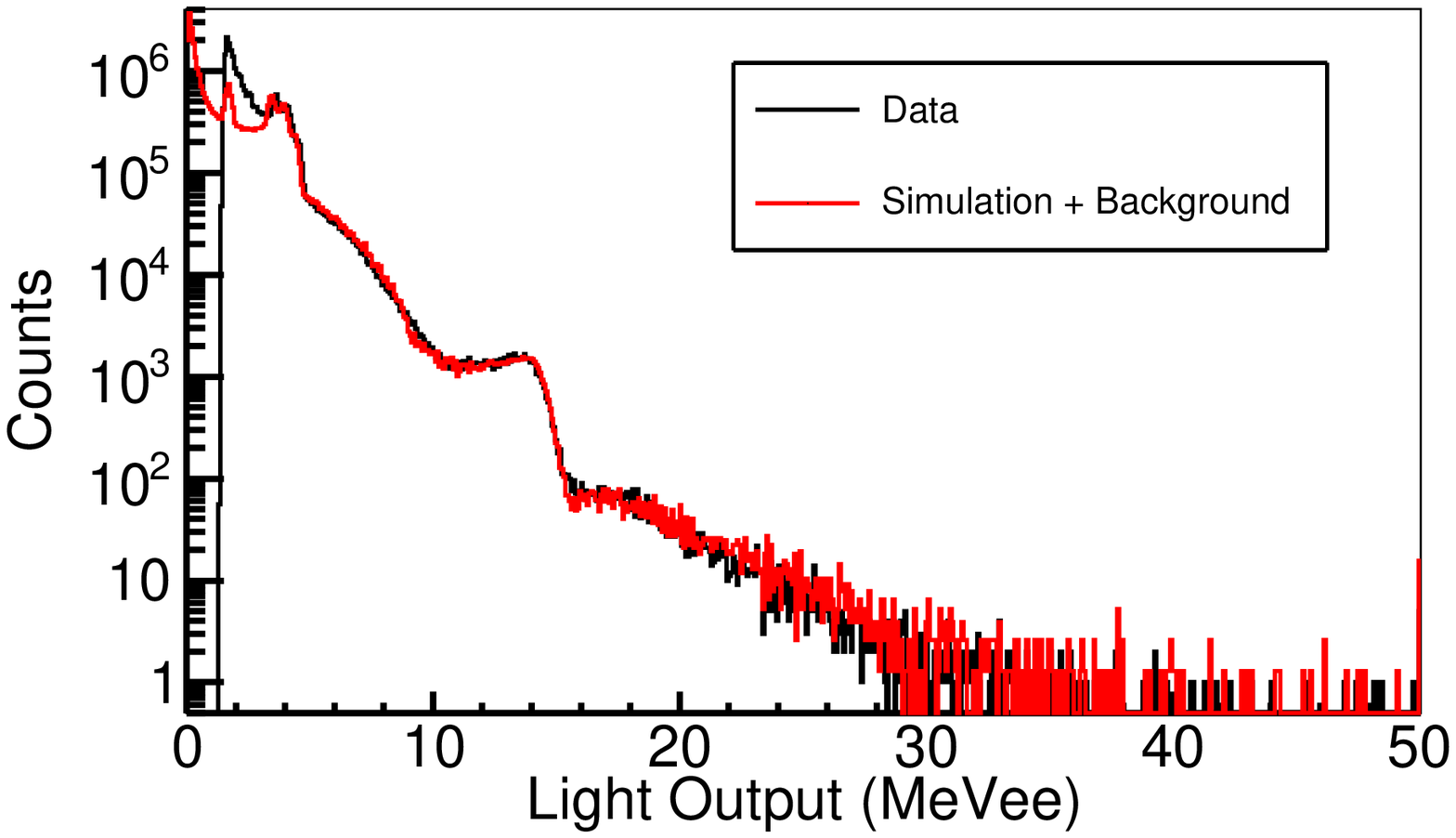}  \\
  \end{tabular}
  \caption{Recorded light-output spectra measured with a (a)~3'' EJ309 detector detector at 90\textdegree~and (b)~1.5'' NaI(Tl) detector at 0\textdegree~for a 25-MeV proton incident onto a thick natural carbon target over a measurement time of 8~hours.}
 \label{fig:fitresult}
\end{figure}

The simulated EJ309 light-output responses to 4.4-MeV gamma rays, 15.1-MeV gamma rays, and bremsstrahlung are scaled and added to the background spectrum such that their sum matches the experimental light-output response. Individual contributions to the simulated spectra, the combined simulated spectra, and the experimental spectra for the EJ309 detector at 90\textdegree~and the NaI(Tl) detector detector at 0\textdegree~are shown in Fig.~\ref{fig:indCont}. In this example, protons ($E_p=25$~MeV) are incident onto a thick natural carbon target over a measurement time of 8~hours.
The final combined fit for one of the experimental configurations for the NaI(Tl) detector and EJ309 detector is shown in Fig.~\ref{fig:fitresult}. The fit spectra show good agreement over the broad light-output range of interest. 

By calculating the area under the simulated detector response for the individual contributions of 15.1- and 4.4-MeV gamma rays and knowing the on-target current used during the experiment and the experimental measurement time, the absolute detection efficiency obtained from the simulation, and assuming an isotropic angular photon distribution, we estimate the production rates of 4.4- and 15.1-MeV gamma rays. Pileup corrections have been applied to the production yields in accordance with the values listed in Table~\ref{table:data_rate}. Due to unidentifiable sources of uncertainty in our measurements such as proton energy and time-varying backgrounds, a conservative approach to estimate the uncertainty was pursued. The bin errors of the measured light-output spectra were varied to achieve a $\chi^2$/NDF=1 for each measurement. The bin errors range from approximately 20 to 40\%. 
\section{Results \& Discussion}
\begin{figure}[]
\vspace{-.5em}
\centering
\includegraphics[width=\linewidth]{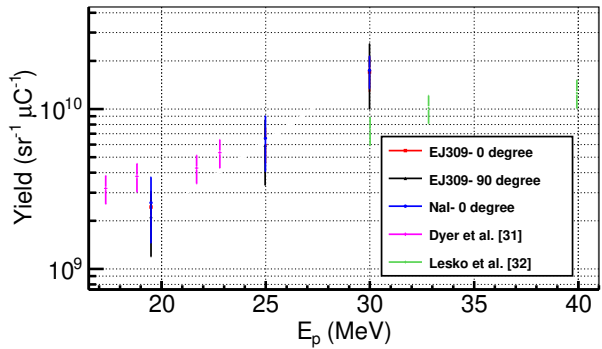}
\caption{Gamma-ray yields at 4.4~MeV for EJ-309 detectors and single NaI(Tl) detector at proton energies of 19.5, 25, and 30~MeV.}\label{Fig:yields44}
\vspace{-.5em}
\end{figure}

\begin{figure}[]
\vspace{-.5em}
\centering
\includegraphics[width=\linewidth]{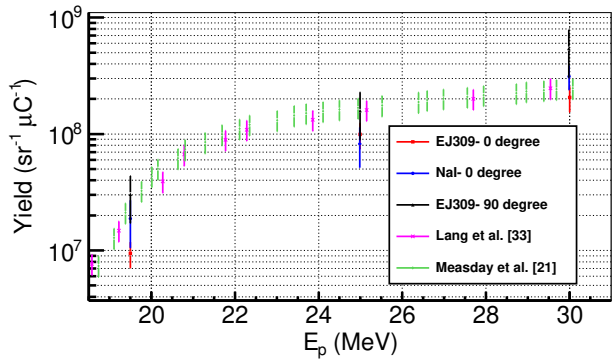}
\caption{15.1-MeV gamma-ray yields for both EJ-309 detectors and single NaI(Tl) detector for proton energies of 19.5, 25, and 30~MeV. }\label{Fig:yields15}
\vspace{-.5em}
\end{figure}

The results show an expected trend: as the energy of the protons is increased, the yield of high-energy gamma rays increases. The 15.1-MeV gamma-ray yield at 0\textdegree~from $^{12}$C(p,p')$^{12}$C reaction is a factor of 1.3, 14.8, and 55.5 (sr$^{-1}$ \textmu C$^{-1}$) greater at proton energies of 19.5, 25, and 30~MeV, respectively, than the yield of 15.1-MeV gamma rays from $^{11}$B(d,n)$^{12}$C reaction at 0\textdegree~angle with 3-MeV deuteron energy~\cite{Rose2016}. The increase in the 15.1-MeV gamma-ray yield measured at 90\textdegree~is even greater compared to the $^{11}$B(d,n)$^{12}$C reaction. A summary of 4.4- and 15.1-MeV gamma-ray yield is listed in Table~\ref{table:44} and~\ref{table:15} with the corresponding values plotted in Fig.~\ref{Fig:yields44} and Fig.~\ref{Fig:yields15}, respectively. 4.4-MeV and 15.1-MeV gamma-ray yields per sr \textmu C from were calculated by 
\begin{equation}
Y(E) = \frac{npN_{\textrm{A}}}{M} \int_{E_0}^{E_1} \frac{\sigma(E)}{T(E)}  dE,
\end{equation}
where \textit{n} is the number of protons, \textit{p} the enrichment of the $^{12}$C sample, \textit{M} is the molar mass, \textit{E} the proton energy, $\sigma$ the measured production cross section~\cite{dyer1981cross,lesko1988measurements,lang1987cross,measday:1963}, and \textit{T} the tabulated stopping power~\cite{pstar}. The estimated yields are shown in Fig.~\ref{Fig:yields44} and Fig.~\ref{Fig:yields15} assuming a 20\% uncertainty. The gamma-ray yields are calculated by integration from the incident energy of the proton to the reaction threshold of its respective gamma-ray energy. There is good agreement between the measured gamma-ray yields at 4.4 and 15.1~MeV and previous work~\cite{dyer1981cross,lesko1988measurements,lang1987cross}. The differences between our experimental results and calculated yields based on previously reported cross sections may be due to the lack of fine-resolution cross-section data and our assumption of isotropic angular distributions.
The threshold for neutron production in the $^{12}$C(p,n) reaction is 19.64~MeV~\cite{rimmer1968resonances}. Neutrons can also be produced below this threshold, for example from isotopic impurities such as $^{13}$C in the target and from energetic gamma rays produced in the target and interacting in the surrounding material. The presence of fast neutrons is confirmed in Fig.~\ref{fig:LOvsPSP}, where proton energy was below the $^{12}$C(p,n) threshold. The results indicate that, as the energy of the protons is increased, the endpoint of the light-output distribution corresponding to neutrons increases, and therefore there is also an increasing trend in the produced neutron energy. We also observe an increase in the total number of produced neutrons with increasing incident proton energy. 
\begin{table}[h!]
		\vspace{-.5em}
		\caption{4.4-MeV gamma-ray yield measured with 3'' EJ-309 detector.  
		}
		\label{table:excite}
		\centering
		\begin{tabular}{ccc}
			\hline
			\hline
			 Proton Energy (MeV) & Position  & Yield (sr$^{-1}$ \textmu C$^{-1}$) \\
			\hline
			\multirow{2}*{19.5} & 0\textdegree  & 2.44$\times10^9$$\pm6.1\times10^8$ \\
			                    & 90\textdegree & 2.12$\times10^9$$\pm9.3\times10^8$\\\hline
			 \multirow{2}*{25}  & 0\textdegree  & 6.07$\times10^9$$\pm1.5\times10^9$ \\
			                    & 90\textdegree & 5.91$\times10^{9}$$\pm2.6\times10^9$\\\hline
			 \multirow{2}*{30}  & 0\textdegree  & 1.69$\times10^{10}$$\pm4.2\times10^9$ \\
			                    & 90\textdegree &1.77$\times10^{10}$$\pm7.8\times10^9$\\\hline\hline
		\end{tabular}	
		\label{table:44}
		\vspace{-.5em}
\end{table}
\begin{table}[h!]
		\vspace{-.5em}
		\caption{15.1-MeV gamma-ray yield measured with 3'' EJ-309 detector.}
		\label{table:excite}
		\centering
		\begin{tabular}{ccc}
			\hline
			\hline
			 Proton Energy (MeV) & Position  & Yield (sr$^{-1}$ \textmu C$^{-1}$) \\
			\hline
			\multirow{2}*{19.5} & 0\textdegree  & 9.38$\times10^6$$\pm2.3\times10^6$ \\
			                    & 90\textdegree & 3.02$\times10^7$$\pm1.3\times10^7$\\\hline
			\multirow{2}*{25}   & 0\textdegree  & 9.91$\times10^7$$\pm2.5\times10^7$ \\
			                    & 90\textdegree & 1.58$\times10^8$$\pm7.0\times10^7$\\\hline
			\multirow{2}*{30}   & 0\textdegree  & 2.06$\times10^8$$\pm5.1\times10^7$ \\
			                    & 90\textdegree & 5.37$\times10^8$$\pm2.4\times10^8$\\\hline\hline
		\end{tabular}	
		\label{table:15}
		\vspace{-.5em}
\end{table}
%
%

The photon and neutron dose equivalents can be obtained from the measured total readings $Q_7$ and $Q_6$ of TLD~700 and TLD~600 and are shown in Table~\ref{table:TLD_reading}. Each reading is averaged from the results of three independent chips and the uncertainty is quoted as the associated standard deviation. Because the photon flux is larger than the neutron flux in the experimental environment and the TLD~700 is much less sensitive to neutrons than TLD 600 (for neutron energy $\lesssim10$~MeV)~\cite{Hsu2008}, the total readings $Q_7$ of TLD~700 are dominated by the photon response, such as $Q_7 \simeq Q_{7,p}$. Therefore, the photon dose listed in Table~\ref{table:TLD_dose} is estimated from Eq.~(\ref{g_calib}) and the 8-hour measurement time. Following the calculation using Eqs.~(\ref{6_g_read}) and~(\ref{6_n_read}), the neutron response $Q_{6,n}$ of TLD~600 can be estimated and is listed in Table~\ref{table:TLD_reading}.

\begin{table}[t]
\vspace{-.5em}
\caption{Results of TLD measurements over a measurement time of 8 hours.}
\label{table:TLD_reading}
\centering
\begin{tabular}{ccccc}
\hline
\hline
Proton & \multirow{2}{*}{Angle}  &  \multirow{2}{*}{\text{TLD 700}}    &  \multicolumn{2}{c}{\multirow{2}{*}{\text{TLD 600}}} \\
energy &   &  & & \\
           (MeV)      &               & $Q_7$(nC)         & $Q_6$ (nC)       & $Q_{6,n}$ (nC)\\
\hline
\multirow{2}*{19.5}   & 0\textdegree  & 8447 $\pm$ 383    & 8731$\pm$1661    & 930$\pm$1698\\
                      & 90\textdegree & 8146 $\pm$ 889    & 9612$\pm$2024    & 2090$\pm$2184\\  
\hline
\multirow{2}*{25}     & 0\textdegree  & 22331 $\pm$ 4520  & 38724$\pm$2384   & 18118$\pm$4811\\
                      & 90\textdegree & 19810 $\pm$ 927   & 34400$\pm$4902   & 16105$\pm$4977\\
\hline
\multirow{2}*{30}     & 0\textdegree  & 35722 $\pm$ 1510  & 151812$\pm$9721  & 118822$\pm$9821\\
                      & 90\textdegree & 33535 $\pm$ 1870  & 117670$\pm$6188  & 86700$\pm$6424\\
\hline
\hline
\end{tabular}	
\label{table:TLD}
\vspace{-.5em}
\end{table} 

\begin{table}[h]
\vspace{.5em}
\caption{Estimated photon and neutron dose distributions from TLD measurements.}
\label{table:TLD_dose}
\centering
\begin{tabular}{cccc}
\hline
\hline
Proton energy          & Angle         &  Photon dose                       &  Neutron dose equivalent\\
        (MeV)          &               &  (\textmu Gy \textmu C$^{-1}$)     &  (\textmu Sv \textmu C$^{-1}$)\\
\hline
\multirow{2}*{19.5}    & 0\textdegree  &  13.69$\pm$0.62                    &  15.83$\pm$28.92\\
                       & 90\textdegree &  13.19$\pm$1.44                    &  35.77$\pm$37.18\\  
\hline
\multirow{2}*{25}      & 0\textdegree  &  42.21$\pm$8.56                    &  359.83$\pm$95.56\\
                       & 90\textdegree &  37.45$\pm$1.76                    &  319.83$\pm$98.85\\
\hline
\multirow{2}*{30}     & 0\textdegree   &  61.39$\pm$2.6                     &  2145.27$\pm$195.04\\
                      & 90\textdegree  &  45.66$\pm$2.53                    &  1565.31$\pm$127.58\\
\hline
\hline
\end{tabular}	
\label{table:TLD}
\vspace{-.5em}
\end{table} 

When $E_p=19.5$~MeV, TLD~600 results indicate that the neutron flux is much lower than the photon flux, so that the neutron reading $Q_{6,n}$ is represents a minor component of the $Q_{6}$ response. This low neutron flux is near the detection limit of TLD~600, resulting in the significant uncertainties of the measured $Q_{6}$ and the estimated $Q_{6,n}$. When the proton energy is increased to $E_p=25$ and 30~MeV, $Q_7$ and $Q_6$ readings also increase as a result of the increased photon and neutron flux. Note that for TLD~600, the neutron reading $Q_{6,n}$ becomes comparable to the corresponding photon reading $Q_{6,p}=Q_{6}- Q_{6,n}$ for $E_p=25$~MeV, while $Q_{6,n}$ can greatly surpass $Q_{6,p}$ with $E_p=30$~MeV. The neutron dose equivalent in Table~\ref{table:TLD_dose} was evaluated using Eq.~(\ref{n_H_calib}). The TLD data in Tables~\ref{table:TLD_reading} and~\ref{table:TLD_dose} indicate a two orders of magnitude increase of the neutron dose equivalent at 0$^{\circ}$ when the proton energy is increased from 19.5 to 30~MeV; a moderate five-fold enhancement is obtained for the photon dose over the same proton energy range.
\begin{figure}
  \centering
  \begin{tabular}{@{}p{1\linewidth}@{}p{1\linewidth}@{}}
    \mysubfigimg[width=\linewidth]{(a)}{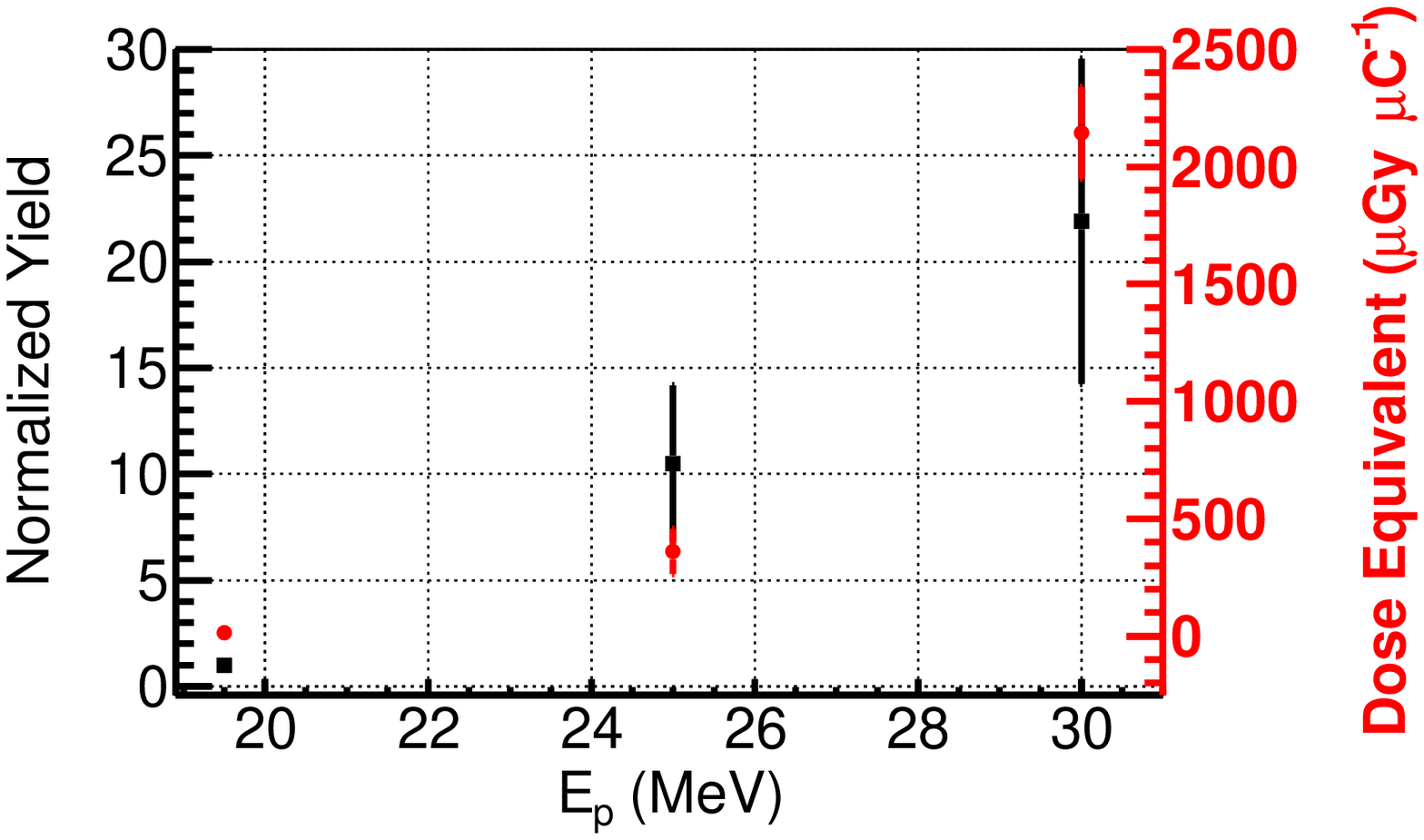} \\
    \mysubfigimg[width=\linewidth]{(b)}{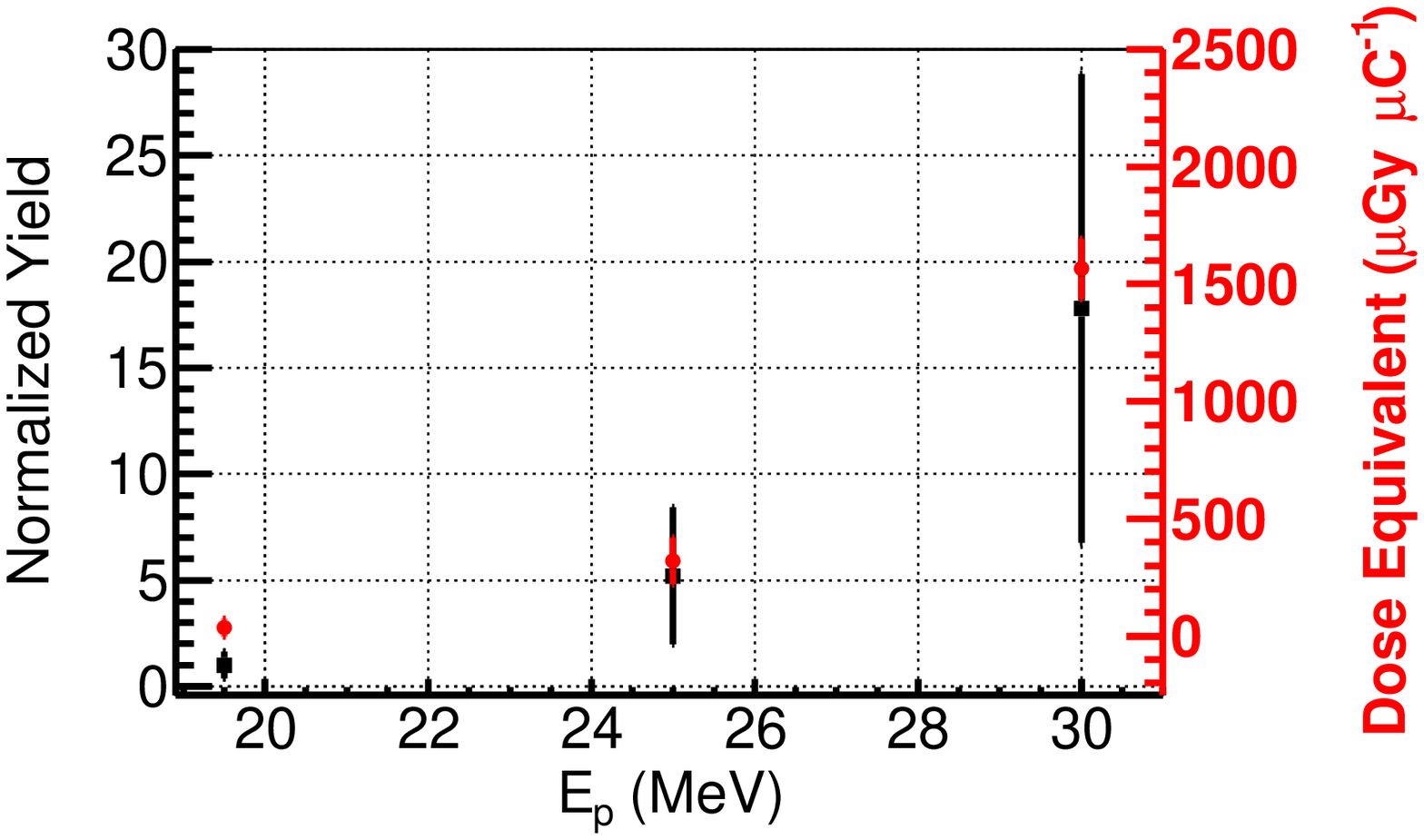}  \\
  \end{tabular}
  \caption{15.1-MeV gamma-ray yield normalized to the yield at E$_p$=19.5~MeV and neutron dose equivalent for incident proton energies of 19.5, 25, and 30~MeV. Results from EJ309 detectors and TLDs are shown at (a)~0\textdegree~and (b)~90\textdegree.}
 \vspace{-1.em}
 \label{fig:pChange1}
\end{figure}
\section{Conclusions}
Shipping manifests reported in Ref.~\cite{descalle2006analysis} show that containers containing high-\textit{Z} material account for less than 0.01\% of cargo. The detection criteria for a high-\textit{Z}-based active interrogation system would have to be carefully selected to ensure an acceptable false-positive rate. A practical system could incorporate a combination of photon and/or neutron radiographic signatures~\cite{nattress2018discriminating} as well as fission signatures such as delayed neutrons~\cite{Mayer2016} to maximize the probability of detection. A comprehensive study is recommended to develop a detailed design and concept of operations for such system, but is beyond the scope of this paper.  

Modest size cyclotrons capable of accelerating protons to 25~MeV with currents up to 400~\textmu A are readily available~\cite{bestC}, and even more compact designs with currents of up to 1~\textmu A are under development~\cite{johnstone15}. Although it is non-portable, the cyclotron described in Ref.~\cite{bestC} could still be installed at a fixed location such as a port of entry. This cyclotron's magnet is 3~m in diameter and weighs 30~tons; its radio-frequency system requires 25~kW for operation. These SWAP parameters are modest in comparison to standard shipyard cranes in both power and weight.

The measured current-normalized thick-target 4.4- and 15.1-MeV gamma-ray yields from $^{12}$C(p,p')$^{12}$C at both 0\textdegree~ and 90\textdegree~are greater than the previously measured $^{11}$B(d,$\gamma$n)$^{12}$C 0\textdegree~yield at the 3-MeV incident deuteron energy for all tested proton energies. The highest 15.1-MeV photon yield was measured at a proton energy of 30~MeV at 90\textdegree, and represents a $\sim$70-fold increase over $^{11}$B(d,$\gamma$n)$^{12}$C at 0\textdegree at the deuteron energy of 3~MeV. The photon dose measurements show increases in photon dose which are nearly commensurate with the increase in the measured 4.4-MeV gamma-ray yield as the proton energy is increased. Figure~\ref{fig:pChange1} shows the gamma-ray yield normalized to the yield at $E_p=19.5$~MeV and neutron dose at three proton energies used in the experiment. Unlike the photon dose, the neutron dose increases at a rate much greater than the 15.1-MeV photon yield as the proton energy is increased in the range of 19.5--30~MeV. There is a two orders of magnitude increase of the neutron dose when the proton energy is raised from 19.5~MeV to 30~MeV, while an approximately five-fold increase is observed for the photon dose over the same proton energy range. Increasing the proton energy up to 30~MeV results in increases of gamma-ray yields, but these are outpaced by the increases in neutron dose.   

The $^{12}$C(p,p')$^{12}$C reaction source could potentially address some of the key technical specifications required for new active interrogation systems~\cite{worlCustoms}. By taking advantage of the emission of the gamma rays into a large solid angle, even multiple cargo scanning streams may be feasible with a single source, thereby increasing the container throughput. With no direct neutron production at proton energies below 19.5~MeV, the $^{12}$C(p,p')$^{12}$C reaction should further reduce the neutron shielding requirements and lower the total radiation dose imparted to cargo or stowaways~\cite{taddeucci2011high}. The reduction in the amount of required shielding would also reduce the overall geometric footprint -- another important consideration in system development. The gamma-ray energies produced by this reaction would provide higher penetration capability than existing bremsstrahlung systems (6- and 9-MeV photon energy endpoints) while enabling robust material discrimination by means of dual-energy photon radiography~\cite{Rose2016, henderson2018experimental}.
%
%
%

\vspace{20pt}
\section*{Acknowledgments}
The authors wish to thank Paul Rose and Anna Erickson for assistance with the CoMPASS data acquisition package. This work was supported by the US Department of Homeland Security [2015-DN-077-ARI096]. The research of J.N. was performed under appointment to the Nuclear Nonproliferation International Safeguards Fellowship Program sponsored by the National Nuclear Security Administration’s Office of International Safeguards (Grant No. NA-241). The research of F.S. was performed under appointment to the Livermore Graduate Scholar Program Fellowship. 
%
\bibliography{INER.bib,rev-tex-custom}
%

\end{document}